\documentclass[12pt, preprint]{aastex} 

\shorttitle{Rapid Migration in a Non-Uniform Nebula}
\shortauthors{Haghighipour $\&$ Boss}

\begin{document}

\title{On Gas-Drag Induced Rapid Migration of Solids 
in a Non-Uniform Solar Nebula}

\author{ Nader Haghighipour and Alan P. Boss}
\affil{Department of Terrestrial Magnetism, Carnegie 
       Institution of Washington, \\ 
       5241 Broad Branch Road, Washington, DC 20015}

\email{nader@dtm.ciw.edu, boss@dtm.ciw.edu}

\begin{abstract}
We study the motions of small solids, ranging from micron-sized
dust grains to meter-sized objects, in the vicinity of local
pressure enhancements of a gaseous nebula. 
Integrating numerically, we show that as a result of 
the combined effect of gas drag and pressure gradients,
solids tend to accumulate at the locations where the
pressure of the gas maximizes. The rate of migration of solids
varies with their sizes and densities and also
with the physical properties of the gas. The results of our
numerical simulations indicate that such migrations are most
rapid for meter-sized objects.
The applicability of the results to the enhancement of the
collision and coagulation of solids and also to the
growth-rate of planetesimals is discussed.
\end{abstract}


\section{Introduction}

We have recently shown that in a rotating, turbulence-free and
isothermal gaseous nebula with a density function that maximizes
locally, the combined effect of gas drag 
and pressure gradients at hydrostatic equilibrium,
causes solids to migrate toward
the locations of local density enhancements
\citep[][hereafter HB03]{Hag03}.
The motions of solids in HB03
were restricted to the midplane of the nebula and
the variation of the gas density along the direction perpendicular 
to the midplane was neglected. The results of
our numerical simulations of the dynamics of micron-sized
dust grains to 100 meter-sized objects indicated that 
the radial motions of such solids could be quite rapid 
when their sizes range from 1 cm to 1 m. 

In this paper, we extend our previous analysis to three-dimensional 
motions and study the radial and vertical migrations of solids 
subject to gas drag and the gravitational
attraction of a solar type star in a 
non-uniform gaseous nebula.

Study of the motions of solids in gaseous mediums has long
been of particular interest to astronomers. 
Because of the great diversity in the applicability
of the results of such studies, in particular, to the
process of formation of planetesimals, 
the literature in this area has become  
so rich that it makes it virtually impossible to cite all the related
articles here. For a short review of the history of such
studies we refer the reader to HB03 and the references
therein. Of direct relevance to this study, however, 
are two classic papers by 
\citet{Ada76} and \citet{Weid77}, an article by 
\citet{Sup00} on the study of the orbital evolution of 
solids in a turbulent protostellar disk with application 
to the formation of icy planetesimals in the outer region 
of the nebula and a recent paper by \citet{Lin02} on the 
radial migration of dust grains in an accreting solar nebula.

Our model nebula is turbulence-free and isothermal. 
We assume that it consists of pure molecular hydrogen
and follows the equation of state of an ideal gas.
An enhancement in the pressure of the gas in such a
nebula appears where the density of the
gas is locally enhanced. To focus attention
on the rate of migration of solids and its dependence
on their sizes and physical properties, 
the density of the nebula is assumed to have an
azimuthally symmetric maximum on the midplane. Along
the vertical axis, the gas density changes in a way that
the vertical component of the pressure gradients balances
the vertical component of the gravitational force of 
the central star. The direction of radial motions of solids
in our model nebula depends on their initial positions 
relative to the location of the maximum gas density and
their vertical migration is always toward the midplane.
\citep{Whip64, Hag03}.

We start by explaining our physical model in $\S$ 2 and
present the equations of motions of solids in $\S$ 3.
In $\S$ 4, we present the results of our numerical 
simulations and in $\S$ 5, we conclude our study by 
reviewing the results and discussing their applications.

\section{The Physical Model}

We consider a rotating, turbulence-free, and isothermal gaseous
nebula of pure molecular hydrogen with a Sun-like star at
its center. At hydrostatic equilibrium and in a cylindrical
coordinate system with the star at its origin and 
the midplane of the nebula as its polar plane, the gravitational
attraction of the star at a point with a position vector
${\bf R}(r,z)$ is related to the pressure gradients
of the gas as
\begin{equation}
r {\omega_g^2} = r {\omega_K^2}\,+\,
{1\over {{\rho_g}({\bf R})}}\,
{{\partial{{\cal P}_g}({\bf R})}\over {\partial r}}
\end{equation}
in $r$-direction and
\begin{equation}
z {\omega_K^2}\,+\,
{1\over {{\rho_g}({\bf R})}}\,
{{\partial{{\cal P}_g}({\bf R})}\over {\partial z}}=0
\end{equation}
along the $z$-axis.
In equations (1) and (2), ${\rho_g}({\bf R})$ and
${{\cal P}_g}({\bf R})$ are the density and pressure of
the gas, respectively, $\omega_g$
is the angular velocity of the gas on a plane parallel to the
midplane at a height $z$, and $\omega_K$ is its corresponding
Keplerian angular velocity given by
\begin{equation}
{\omega_K^2}={{GM}\over {({r^2}+{z^2})^{3/2}}},
\end{equation}
where $G$ is the gravitational
constant and $M$ is the mass of the central star.

The balance of the gravitational attraction of the central
star by the pressure gradients in the vertical direction,
as given by equation (2), determines the distribution
of the gas along the $z$-axis. In the model nebula
considered here, the gas is pure molecular hydrogen
and it, therefore, obeys the equation state of an ideal gas. 
As a result, the pressure and density of the gas are related as
\begin{equation}
{{\cal P}_g}={{{K_{\rm B}}T}\over {m_0}}{\rho_g}({\bf R})\,,
\end{equation}
where $K_{\rm B}$ is the Boltzmann constant, $T$ is the temperature
of the gas and $m_0$ is the molecular mass of hydrogen.
Substituting for ${{\cal P}_g}({\bf R})$ from equation (4)
in equation (2), the density of our model nebula will be given by 
\begin{equation}
{\rho_g}(r,z)\,= \,{\rho_g}(r,0) \>{\rm {Exp}}
\Biggl\{{{GM{m_0}}\over {{K_{\rm B}}T}}\,
\biggl[{1\over {{({r^2}+{z^2})^{1/2}}}}-{1\over r}\biggr]\Biggr\}.
\end{equation}
In equation (5), ${\rho_g}(r,0)$ is the density of the gas 
on the midplane.

As mentioned in $\S$ 1, in order to focus attention
on the dynamics of solids, we consider a nebula with an
azimuthally symmetric density enhancement on its midplane.
Following HB03, we assume
\begin{equation}
{\rho_g}(r,0)\,=\,{\rho_0}\,
{\rm {Exp}}\Biggl[-\beta\Bigl(
{r \over{r_m}}-1 {\Bigr)^2}\Biggr]\,,
\end{equation}
where ${\rho_0}\,,{r_m}$ and $\beta$ are constant quantities.
Figure 1 shows an edge-on view of our model nebula with the line of 
sight perpendicular to 
one of the midplane diameters. The central star in this figure
is solar-type and the gas temperature is 300 K. The values of
${\rho_0}\,,\beta$ and ${r_m}$ are equal to
${10^{-9}}\,{\rm {g\,cm^{-3}}}\,,1$ and 1 AU, respectively.

\section{Equation of Motion}

The equation of motion of an object with mass of unity
in our model nebula is given by
\begin{equation}
{{{d^2}{\bf R}}\over {d{t^2}}}\,=\,
-\,{{GM}\over {({r^2}+{z^2})^{3/2}}}\,{\bf R}\,-\,{{\bf F}_{\rm {drag}}}
\end{equation}
\noindent
where ${\bf F}_{\rm {drag}}$ represents the
drag force of the gas for the unit mass of a solid. 
Following \citet{Sup00} and as presented in HB03,
we write ${\bf F}_{\rm {drag}}$ as
\begin{equation}
{{\bf F}_{\rm {drag}}}\,=\,
{{{\rho_g}({\bf R})}\over {{\rho_p}({a_p}+\lambda)}}\,
\Biggl[\Bigl({\lambda\over {a_p}}\Bigr)\,
{{\bar v}_{\rm {th}}}\,+\,
{3\over 8}\,{C_D}\,{v_{\rm {rel}}}\Biggr]\,
{{\bf V}_{\rm {rel}}}\,.
\end{equation}
\noindent
In equation (8), $a_p$ is the radius of the object, $\rho_p$
denotes its density, ${v_{\rm {rel}}}=|{{\bf V}_{\rm {rel}}}|$
is the magnitude of its velocity relative to the
gas and $\lambda$ is the mean free path of the gas molecules.
For a nebula of pure molecular hydrogen, 
$\lambda\,{\rm {(cm)}}=4.79 \times {10^{-9}}/
{{\rho_g}({\bf R})}$ g cm$^{-3}$ (HB03). The quantity
${{\bar v}_{\rm {th}}}=(8{K_{\rm B}}T/\pi {m_0})^{1/2}$ in equation (8)
is the mean thermal velocity of the gas molecules and
$C_D$ is the drag coefficient whose numerical value is given by
\citep{Whip64, Weid77}
\vskip 3pt
\begin{equation}
{C_D}\,\simeq\,\cases{
24\,{{\rm Re}^{-1}} & if$\quad {\rm Re}<1$ (Stokes' drag) \cr
24\,{{\rm Re}^{-0.6}} & if$\quad1<{\rm Re}<800$\cr
0.44&if$\quad {\rm Re}>800$.\cr}
\end{equation}
\vskip 3pt
\noindent
In equation (9), ${\rm Re}$ is the gas Reynolds number and is equal to 
\begin{equation}
{\rm Re}\,=\,\Bigr({{6\sigma {a_p}} \over 
{{m_0}{{\bar v}_{\rm {th}}}}}\Bigl) 
{\rho_g}({\bf R}) \,{v_{\rm rel}}\,,
\end{equation}
\vskip 2pt
\noindent
where $\sigma=2\times {10^{-15}}\,{\rm {cm}}^2$ 
is the collisional cross section of hydrogen molecules.

The velocity of a solid relative to the gas is equal to
${{\bf V}_{\rm {rel}}}={{\bf V}_p}-{{\bf V}_g}$ where
${\bf V}_p$ and ${\bf V}_g$ represent the velocities of the
solid and gas molecules, respectively.
In the cylindrical coordinate system considered here,
${{\bf V}_{\rm {rel}}}$ has the following components.
\begin{eqnarray}
&\!\!\!\!\!\!\!\!\!\!\!\!\!\!\!\!\!\!\!\!\!\!\!\!\!\!
\!\!\!\!\!\!\!\!\!\!\!\!\!\!\!\!\!\!\!\!\!\!\!\!\!\!
\!\!\!\!\!\!\!\!\!\!\!\!\!\!\!\!\!\!\!\!\!\!\!\!\!\!
\!\!\!\!\!\!\!\!\!\!\!\!\!\!\!\!\!\!\!\!\!\!\!\!\!\!
{({{\bf V}_{\rm {rel}}})_r}\,=\,{\dot r}\,,\\
&\!\!\!\!\!\!\!\!\!\!\!\!\!\!\!\!\!\!\!\!\!\!\!\!\!\!
\!\!\!\!\!\!\!\!\!\!\!\!\!\!\!\!\!\!\!\!\!\!\!\!\!\!
\!\!\!\!\!\!\!\!\!\!\!\!\!\!\!\!\!\!\!\!\!\!\!\!\!\!
\!\!\!\!\!\!\!\!\!\!\!\!\!\!\!\!\!\!\!\!\!\!\!\!\!\!
{({{\bf V}_{\rm {rel}}})_z}\,=\,{\dot z}\,,\\
&\!\!\!\!\!\!\!\!\!\!\!\!\!\!\!\!\!\!\!\!\!\!\!\!\!\!
\!\!\!\!\!\!\!\!\!\!\!\!\!\!\!\!\!\!\!\!\!\!\!\!\!\!
\!\!\!\!\!\!\!\!\!\!\!\!\!\!\!\!\!\!\!\!\!\!\!\!\!\!
\!\!\!\!\!
{({{\bf V}_{\rm {rel}}})_\varphi}\,=\,
r({\dot \varphi}\,-\,{\omega_g})\,.
\end{eqnarray}
\noindent
where we have assumed that the gas molecules do not move along the
$z$-axis. It is important to mention that it is the $r$ and $z$ components
of ${{\bf V}_{\rm {rel}}}$ that play important roles in the
radial and vertical migrations
of solids. They also dominate the magnitude of  ${{\bf V}_{\rm {rel}}}$
when the object is small.
The transverse component of the relative
velocity, ${({{\bf V}_{\rm {rel}}})_\varphi}$,
will have a dominating effect in the magnitude of  ${{\bf V}_{\rm {rel}}}$
for larger objects and it plays an important role 
in changing the angular momenta of such solids which can
have significant contributions when accretion is considered.

From equations (11) to (13), the equation of motion of a 
solid with a unit mass can be written as
\begin{eqnarray}
&\!\!\!\!\!\!\!\!\!\!\!\!\!\!\!\!\!\!\!\!\!\!\!\!\!\!
\!\!\!\!\!\!\!\!\!\!\!\!\!\!\!\!\!\!\!\!\!\!\!\!\!\!
\!\!\!\!\!\!\!\!\!\!\!\!\!\!\!\!\!\!\!\!\!\!\!\!\!\!
{{\hat P}_r}={\dot {\hat r}},\\
&\!\!\!\!\!\!\!\!\!\!\!\!\!\!\!\!\!\!\!\!\!\!\!\!\!\!
\!\!\!\!\!\!\!\!\!\!\!\!\!\!\!\!\!\!\!\!\!\!\!\!\!\!
\!\!\!\!\!\!\!\!\!\!\!\!\!\!\!\!\!\!\!\!\!\!\!\!\!\!
{{\hat P}_z}={\dot {\hat z}},\\
&\!\!\!\!\!\!\!\!\!\!\!\!\!\!\!\!\!\!\!\!\!\!\!\!\!\!
\!\!\!\!\!\!\!\!\!\!\!\!\!\!\!\!\!\!\!\!\!\!\!\!\!\!
\!\!\!\!\!\!\!\!\!\!\!\!\!\!\!\!\!\!\!\!
{{\hat P}_\varphi}={{\hat r}^2}\,{\dot {\hat \varphi}},\\
&\!\!\!\!\!\!\!\!\!\!\!\!\!\!\!\!\!\!\!\!\!\!\!\!\!\!
\!\!\!\!\!\!\!\!\!\!\!\!\!\!\!\!\!\!\!\!\!
\!\!\!\!\!\!\!\!\!\!\!\!\!\!\!\!\!\!\!\!\!
\!\!\!\!\!\!\!\!\!\!\!\!\!\!\!\!\!\!\!
{{\dot {\hat P}}_r}=& 
\!\!\!\!\!\!\!\!\!\!\!\!\!\!\!\!\!\!\!
\!\!\!\!\!\!\!\!\!\!\!\!\!\!\!\!\!
{1\over {{\hat r}^3}}{{\hat P}_\varphi^2}-
{{\hat r}\over {({{\hat r}^2}+{{\hat z}^2})^{3/2}}}-
{({{\hat F}_{\rm {drag}}})_r},\\
&\!\!\!\!\!\!\!\!\!\!\!\!\!\!\!\!\!\!\!\!\!\!\!\!\!\!
\!\!\!\!\!\!\!\!\!\!
\!\!\!\!\!\!\!\!\!\!\!\!\!\!\!\!\!\!\!\!\!\!\!\!\!\!
\!\!\!\!\!\!\!\!\!\!\!\!\!\!\!\!\!\!\!\!\!\!\!\!\!\!
{{\dot {\hat P}}_z}=&
\!\!\!\!\!\!\!\!\!\!\!\!\!\!\!\!\!\!\!\!\!\!\!\!\!\!
\!\!\!\!\!\!\!\!\!\!\!
-{{\hat z}\over {({{\hat r}^2}+{{\hat z}^2})^{3/2}}}
-{({{\hat F}_{\rm {drag}}})_z},\\
&\!\!\!\!\!\!\!\!\!\!\!\!\!\!\!\!\!\!\!\!\!\!\!\!\!\!
\!\!\!\!\!\!\!\!\!\!\!\!\!\!\!\!\!\!\!\!\!\!\!\!\!\!
\!\!\!\!\!
{{\dot {\hat P}}_\varphi}=
-\,{\hat r}{({{\hat F}_{\rm {drag}}})_\varphi}\,.
\end{eqnarray}
\noindent
In equations (14) to (19), the hat sign indicates a dimensionless
quantity. We have followed our convention presented in HB03
and have chosen two quantities
$r_0$ and $t_0$ to represent the units of length and time,
respectively, in such a way that $GM{t_0^2}{r_0^{-3}}=1$. The
quantities ${{\hat P}_r}\,,{{\hat P}_z},$ and ${{\hat P}_\varphi}$
are the dimensionless radial, vertical and angular momenta of
the object, 
$(r,z)\,=\,{r_0}({\hat r},{\hat z})\,,
{{\hat {\bf F}}_{\rm {drag}}}= {{t_0^2}{{\bf F}_{\rm {drag}}}/{r_0}}$
and $\varphi={\hat \varphi}$.

\section{Numerical Results}

To start integrating equations (14) to (19), numerically, 
we take the mass of the central star to be equal to one solar
mass and we choose, $\beta =1$ and ${\rho_0}={10^{-9}}$g cm$^{-3}$. 
As in HB03, the azimuthal symmetry of the gas density
on the midplane of our model nebula allows us to set 
${r_0}={r_m}=1$ AU which implies
$t_0$=0.16 years.

An important quantity that has to be determined at the
beginning of our integrations is the drag
coefficient, $C_D$. As mentioned in the last section, this
quantity is a function of the Reynolds number of the
gas as given by equation (9).
At the beginning of the integration,
the initial functional form of $C_D$ is 
determined by calculating the initial value of
${\rm Re}$. During the course of integration,
the value of ${\rm Re}$
is constantly monitored. Once the range of this quantity
changes, the integration is continued by adjusting the
drag coefficient to its new functional form.

We integrated equations (14) to (19)
for different values of solids' radii and densities and also
for different values of the gas temperature. The objects
were initially placed on circular orbits at different radial
positions. The initial height $z$ of an object
was taken to be one-tenth of its initial radial distance
and ${\hat \varphi}(t=0)=0$.
The initial velocity along the $z$-axis was set equal to zero. 

\subsection{Radial Migration}

Figures 2 and 3 show the radial migration of objects with radii
ranging from 1 micron to 1 meter. The densities of the objects
are equal to 2 g cm$^{-3}$ and the temperature of the
gas is 300 K. Each object is
once placed at (2,0.2) AU with a radial
distance larger than the radial location of the density enhancement
on the midplane (i.e., $r=1$ AU), and is once
placed at a closer radial distance at (0.25,0.025) AU.
As expected, all objects migrate radially in/out toward $r=1$ AU
(HB03). Smaller objects show the tendency of staying with
the gas and undergo very slow radial migrations.
Larger objects, on the other hand, show more rapid motions.
As in the case of the two-dimensional model (HB03), a 
meter-sized object undergoes the most rapid radial migration. 

Figures 2 and 3 also show that the rate of radial migration
is greater for inward motions. A result that was 
also observed in two-dimensional simulations of HB03. This
can be attributed to the fact that the rate of change
of angular momentum for an object beyond the radial location
of the density enhancement is larger than that of 
an object in a closer distance to the central star.
As a result, farther objects lose angular momentum
faster, hence, their more rapid radial migration 
(see HB03 for more detailed analysis).

As mentioned before, as a result of the combined effect
of gas drag and pressure gradients, it is
expected that objects at radial distances larger than $r_m$
to undergo inward radial migrations.
A closer investigation of the results shown in Figures 2 and 3, however,  
indicates that at the beginning of their radial motions,
objects at $r=2$ AU have, briefly, undergone outward migrations.
Figure 4 shows this for a millimeter- and a centimeter-sized particle.
Such a behavior can be understood by studying the
structure of our model nebula. As given by equation (5), 
the maxima of the gas density on planes parallel to the midplane 
at different heights $z$ will appear at different
radial distances. Figure 5 shows this for
$z$ ranging from zero to 0.2 AU. In the top graph of this figure
we have plotted the density of the gas as a function of $r$ 
for constant values of $z$. As shown here, the radius of the
circular enhancement of the nebula increases from $r=1$ AU
on the midplane to larger values at higher vertical distances. 
The middle and the bottom graphs of Figure 5 show the
gas density enhancement at $z=$ 0.15 and 0.2 AU, respectively. 
When an object is initially at a radial
distance beyond 1 AU, depending on its height from the midplane,
it may be inside the circle of the local density enhancement
at that height. In such cases, the object starts its radial
motion by an outward migration while descending toward the
midplane. Once it passes the height where the radial distance
of the local maximum density is smaller than its instantaneous
radial position, the radial motion of the object 
reverses and it starts its inward migration. 

We also numerically integrated equations (14) to (19) for
different values of a solid's density.
Figure 6 shows the radial migrations of 
1 cm and 1 m solids for different values of their densities. 
The temperature of the gas is 300 K.
Similar to the results of two-dimensional simulations
(HB03), increasing the solid's density results in a more 
rapid radial migration for a centimeter-sized or smaller particle.
However, it has a reverse effect on the motions of
meter-sized and larger objects.
Such different responses to the change of the density of a solid
are the consequences of the contribution of gas drag in 
changing the rates of its radial migration.  
Recall that, as mentioned in $\S$ 3, the radial component
of the velocity of an object relative to the gas is the dominant
factor in its radial motion. For a centimeter-sized or smaller object, 
the magnitude of this component, given by equations (11) and (14), 
also dominates the magnitude
of the relative velocity of the object. Increasing
the density of an object while keeping its radius constant
results in decreasing the magnitude of the drag force
${{\bf F}_{\rm {drag}}}$ given by equation (8). 
Such a decrease in the magnitude of ${{\bf F}_{\rm {drag}}}$, in turn,
causes the rate of change of the radial component
of the relative velocity of the object (equation [17]) to increase,
and as a result the object migrates more rapidly.

The situation is not the same for meter-sized and larger objects.
In such cases, it is the transverse component of 
${{\bf V}_{\rm {rel}}}$ (equation [13])
that dominates its magnitude.
That is, as opposed to small particles,
the motion of a meter-sized or larger object is
dominated by the rate of the change of its angular momentum. 
Because the magnitude of ${{\bf F}_{\rm {drag}}}$ decreases by
increasing the solid's density, from equation (19), 
a larger value for $\rho_p$
results in a smaller rate for the decrease/increase of the 
angular momentum of the object. In other words, an increase of 
the solid's density will increase its reluctance in loosing/gaining
angular momentum and hence a decrease in its rate of 
inward/outward radial migrations.

Numerical integrations were also carried out for different values
of the gas temperature. Figure 7 shows radial migrations of
a centimeter- and a meter-sized object with a density of 2 g cm$^{-3}$
for four values of the gas temperature. As shown here, increasing the
temperature of the gas results in increasing the rate of 
radial migration. This is an expected result which is
the consequence of an increase in the radial component of the
pressure gradients of the
gas. Figure 8 shows this concept in more detail. In this figure,
we have plotted the density of the gas as a function of $r$ on
a plane parallel to the midplane at $z=0.2$ AU. As a result of 
increasing the temperature of the gas from 50 K to 1000 K, the
maximum value of the gas density on this plane 
is increased by three orders of
magnitude. Such an increase implies an increase in the 
radial component of the pressure
gradients of the gas which in turn results in more rapid radial
migration. 

From equation (5), the radial position of
the maximum of the gas density at a certain height $z$
approaches smaller values by increasing the gas temperature.
Figure 8 shows this for $z=$0.2 AU. For an object at this height
and at an initial radial position of $r=2$ AU,
such a decrease in 
the radius of the local density enhancement
causes the distance of the initial outward migration 
of the object to become smaller (Figure 7). 
For the temperature of 1000 K,  
the radial location of the maximum gas density at $z=0.2$ AU 
becomes smaller than 2 AU and as a result, 
an object at (2,0.2) AU does not
undergo an initial outward migration.

\subsection{Vertical Migration}

While the combined effect of pressure gradients
and the drag force of the gas causes solids to
radially migrate toward the location of local
density enhancement, 
the vertical component of the gravitational force of the
central star attracts solids
toward the midplane of the nebula.
Figure 9 shows the vertical motion of solids with different sizes.
The dependence of the rate of vertical migration on the size
of an object can clearly be seen from this figure. 
As shown here, except for
when the objects are in the vicinity of the midplane, the rates of their
vertical migrations increase by increasing their radii.
In the vicinity of the midplane, however, while 10 centimeter-sized
and smaller objects continue their smooth descent, the 1 meter-sized
object undergoes an overshoot. Such an overshoot and its 
corresponding damped oscillatory motion are more pronounced
when the density of the object is increased. Figure 10 shows
the vertical migrations of a 1 centimeter-sized and a 1 meter-sized
object for different values of their densities. As shown here,
the rates of vertical descent of both objects increase by increasing
their densities. Once in the vicinity of the midplane,  
the meter-sized object undergoes a damped oscillatory motion
whose amplitude increases by increasing the solid's density.
Figure 11 shows the three-dimensional paths of these two objects.

The dynamical behavior of an object along the $z$-axis and
in the vicinity of the midplane
can be explained by studying equation (18) in more detail.
In this equation, the drag coefficient $C_{\rm D}$ is the 
factor that determines the functional form of 
${{\dot {\hat P}}_z}$. Substituting for the drag force of 
the gas from equation (8) and for the $z$-component of the 
relative velocity of an object with respect
to the gas from equation (12), equation (18) can be written as
\vskip 5pt
\begin{equation}
{{\dot {\hat P}}_z}= -
{{\hat z}\over {({{\hat r}^2}+{{\hat z}^2})^{3/2}}}-
{{{{\hat \rho}_g}({\bf R})}\over {{{\hat \rho}_p}
({{\hat a}_p}+{\hat \lambda})}}\,
\Biggl[\Bigl({{\hat \lambda}\over {{\hat a}_p}}\Bigr)\,
{{\hat {\bar v}}_{\rm {th}}}\,+\,
{3\over 8}\,{C_D}\,{{\hat v}_{\rm {rel}}}\Biggr]
{{\hat P}_z}\,.
\end{equation}
\vskip 10pt
\noindent
In the following we study equation (20) for 
${\rm Re}<1$ and $1<{\rm Re}<800$. The case of
${\rm Re}>800$ is not studied here since it corresponds 
to the motions of objects
with sizes equal or greater than 100 m. As shown in HB03, 
the rates of migrations of such objects due to gas drag
and pressure gradient decrease by increasing their
sizes. In this paper, we focus our study on rapid migration
of small solids.

\subsubsection{${C_D}=24\, {\rm Re}^{-1}$}

Replacing ${t_0}, {K_{\rm B}}, {m_0}$ and $\sigma$ by
their numerical values and considering spherical solids with
masses of $(4/3)\pi {a_p^3}{\rho_p}$, equation
(20) is written as
\begin{equation}
{{\dot {\hat P}}_z}\simeq -
{{\hat z}\over {({{\hat r}^2}+{{\hat z}^2})^{3/2}}}-
{3.73 \over {{a_p}{\rm (cm)}\,
{\rho_p}({\rm {g\,cm^{-3}}})}\,
[{a_p}{\rm (cm)}+{\lambda {\rm (cm)}}]}\,
{T^{1/2}}\,{{\hat P}_z},
\end{equation}
where $T$ is in Kelvin. In the vicinity of the midplane,
${\hat z}<<{\hat r}$ and equation (21) can be simplified to
\begin{equation}
{\ddot {\hat z}}+
{3.73 \over {{a_p}{\rm (cm)}\,
{\rho_p}({\rm {g\,cm^{-3}}})}\,
[{a_p}{\rm (cm)}+{\lambda {\rm (cm)}}]}\,
{T^{1/2}}\,{\dot{\hat z}}+
{{\hat z}\over {{\hat r}^3}} \simeq 0.
\end{equation}

Equation (22) resembles the equation of a damped 
harmonic motion with a dimensionless frequency of
\begin{equation}
{{\hat \omega}^2}={{\hat r}^{-3}}\,-\,
3.48\,T\,{\Bigl\{{a_p}{\rm (cm)}\,
{\rho_p}({\rm {g\,cm^{-3}}})\,
[{a_p}{\rm (cm)}+{\lambda {\rm (cm)}}]\Bigr\}^{-2}}\,.
\end{equation}
Equation (23) indicates that ${\hat \omega}^2$ is a function
of the solid's radius and density, the gas temperature and also
the mean free path of the gas molecules, $\lambda$. We recall
that $\lambda$ varies with the gas density as 
mentioned in $\S$3.

At any location in the vicinity of the midplane,
the type of the motion of an object depends on
the sign of ${\hat \omega}^2$ at that location.
The left column of Figure 12 shows the absolute value of the
second term of equation (23) for a centimeter-sized object
initially at (1.25,0.125) AU. We chose ${a_p}=1$ cm
as a representative of solids with ${\rm Re}<1$
since the results of our numerical 
integrations indicate that the first sign of
a Reynolds number larger than unity appears for an object
with a radius of 10 cm. Since for an object at
(1.25,0.125) AU, ${\hat r}^{-3}$ varies between 0.512 and 1,
the graphs of ${{\hat r}^2}-{{\hat \omega}^2}$,  
as shown in Figure 12, indicate that for a centimeter-sized
object, ${\hat \omega}^2$ is negative. A negative value for  
${\hat \omega}^2$ is an indication of an overdamped motion.
Increasing the density of the solid
while keeping the gas temperature constant, ${\hat \omega}^2$
becomes larger and approaches zero (critically damped) 
indicating that the overdamped
nature of the vertical motion of the particle weakens and it may
even change into an underdamped motion where  ${\hat \omega}^2$
is positive.

From equation (23),
the overdamping nature of the vertical motion 
becomes even more pronounced when the radius of the object
is smaller than 1 cm. That means, the rate of vertical
migration of a solid decreases by decreasing its size
(Figure 9).

The left column of 
Figure 12 also shows that for a constant value of the
solid's density, increasing the gas temperature results in
lower values for  ${\hat \omega}^2$. That means, at higher
temperatures, the overdamping of the vertical motion of the
object becomes more pronounced (top graph of Figure 13). 
This can be explained noting that from equation (5)
and as shown in Figure 15, at any radial distance $r$,
the vertical distribution of the gas maximizes at $z=0$
(i.e., on the midplane). As shown in Figure 14,
the width of the gas distribution function
along the $z$-axis increases by increasing the gas temperature. 
As a result of this broadening, the resistive effect of gas drag
on the vertical motion of solids appear in a larger range
of variable $z$ which results in a longer time for descending toward the
midplane.

\subsubsection{${C_D}=24\, {\rm Re}^{-0.6}$}

In this case, equation (20) can be written as
\vskip 1pt
\begin{eqnarray}
&\!\!\!\!\!\!\!\!\!\!\!\!\!\!\!\!\!\!\!\!\!
\!\!\!\!\!\!\!\!\!\!\!\!\!\!\!\!\!\!\!\!\!
\!\!\!\!\!\!\!\!\!\!\!\!\!\!\!\!\!\!\!\!\!
\!\!\!\!\!\!\!\!\!\!\!\!\!\!\!\!\!\!\!\!\!
{{\dot {\hat P}}_z}\simeq -
{{\hat z}\over {({{\hat r}^2}+{{\hat z}^2})^{3/2}}}-
{2.44 \over {{a_p}{\rm (cm)}\,
{\rho_p}({\rm {g\,cm^{-3}}})}\,
[{a_p}{\rm (cm)}+{\lambda {\rm (cm)}}]}\,
{T^{1/2}}\,{{\hat P}_z} \nonumber\\
&\Biggl\{1 + 8.56\,{[{a_p}({\rm {cm}})]^{0.4}}\,{T^{-0.2}}
{\biggl[{{\hat P}_r^2}+{{\hat P}_z^2}+
\biggl({{{\hat P}_\varphi}\over {\hat r}} - 
{\sqrt {{1\over {\hat r}} - {9.29 \times 10^{-6}}\,T\, 
{\hat r}({\hat r}-1)}\,\biggr)^2}\,\biggr]^{0.2}}\nonumber\\
&\qquad\qquad\qquad\qquad\qquad\qquad\qquad
{\rm {Exp}}\Bigl\{86098\>{T^{-1}}
\biggl[{1\over {{({{\hat r}^2}+{{\hat z}^2})^{1/2}}}}-
{1\over {\hat r}}\biggr]-0.4{({\hat r}-1)^2}\Bigr\}\Biggr\}.
\end{eqnarray}
\vskip 5pt
\noindent
Similar to the previous case, in the vicinity
of the midplane, equation (24) can be simplified to
\begin{eqnarray}
&\!\!\!\!\!\!\!\!\!\!\!\!\!\!\!\!\!\!\!\!\!
\!\!\!\!\!\!\!\!\!\!\!\!\!\!\!\!\!\!\!\!\!
\!\!\!\!\!\!\!\!\!\!\!\!\!\!\!\!\!\!\!\!\!
\!\!\!\!\!\!\!\!\!\!\!\!\!\!\!\!\!\!\!\!\!
{\ddot {\hat z}}+
{2.44 \over {{a_p}{\rm (cm)}\,
{\rho_p}({\rm {g\,cm^{-3}}})}\,
[{a_p}{\rm (cm)}+{\lambda {\rm (cm)}}]}\,
{T^{1/2}}\,{\dot{\hat z}}
\Biggl\{1 + 8.56\, {[{a_p}({\rm {cm}})]^{0.4}}\,{T^{-0.2}}\nonumber\\
&\qquad
{\Biggl[{{\dot{\hat r}}^2}+{{\dot{\hat z}}^2}+
\biggl({\hat r}{\dot{\hat \varphi}} - 
{\sqrt {{1\over {\hat r}} - {9.29 \times 10^{-6}}\,T\, 
{\hat r}({\hat r}-1)}\,\biggr)^2}\,\Biggr]^{0.2}}
{e^{-0.4{({\hat r}-1)^2}}}\Biggr\}+
{{\hat z}\over {{\hat r}^3}} \simeq 0.
\end{eqnarray}
\vskip 2pt
\noindent
For any value of $\hat r$, equation (25) resembles the equation of 
motion of a damped oscillator
with a dimensionless frequency of
\begin{eqnarray}
&\!\!\!\!\!\!\!\!\!\!\!\!\!\!\!\!\!\!\!\!\!\!\!\!\!\!\!
\!\!\!\!\!\!\!\!\!\!\!\!\!\!\!\!\!\!\!\!\!\!\!\!\!\!\!
\!\!\!\!\!\!\!\!\!\!\!\!\!\!\!\!\!\!
{{\hat \omega}^2}={{\hat r}^{-3}}\,-\,
1.49\,T\,{\Bigl\{{a_p}{\rm (cm)}\,
{\rho_p}({\rm {g\,cm^{-3}}})\,
[{a_p}{\rm (cm)}+{\lambda {\rm (cm)}}]\Bigr\}^{-2}}\nonumber\\
&\!\!\!\!\!\!
{\Biggl\{1 + 8.56 \,{[{a_p}({\rm {cm}})]^{0.4}}\,{T^{-0.2}}
{\Biggl[{{\dot{\hat r}}^2}\!+\!{{\dot{\hat z}}^2}\!+\!
\biggl({\hat r}{\dot{\hat \varphi}}\!-\! 
{\sqrt {{1\over {\hat r}} - 
{9.29 \times 10^{-6}} T 
{\hat r}({\hat r}-1)}\,\biggr)^2}\,\Biggr]^{0.2}}\!\!\!\!
{e^{-0.4{({\hat r}-1)^2}}}\Biggr\}^2}\!\!\!.
\end{eqnarray}
\vskip 5pt
\noindent
The right column of Figure 12 shows the absolute value of the second
term of this equation for different values
of the density of a meter-sized object initially
at (1.25,0.125) AU. The radial migration of 
such an object starts at ${\hat r}=1.25$ and terminates at 
${\hat r}=1$, implying that the first term of equation 
(26) will vary between 0.512 and 1.
As shown in Figure 12, at a constant temperature and for
a given value of $\hat r$, the contribution
of the second term of equation (26) is quite small.
As a result, the instantaneous dimensionless frequency 
${\hat \omega}^2$ stays positive for the entire
time of the solid's migration indicating that the motion
of such an object, in both sides of the midplane, is
an underdamped oscillatory motion. 

Figure 12 also shows that
the numerical contribution of the second term of equation (26)
decreases by increasing the solid's density at a constant
temperature. This results in larger frequencies for denser
objects which in turn, results in more
oscillations (Figure 10). One can also see from Figure 12 that
for a constant value of the solid's density, increasing the
gas temperature will increase the numerical value of the
second term of equation (26). This results in a smaller
frequency and less oscillatory behavior (bottom graph of Figure 13). 
It is necessary to mention that the irregular shape of the
second term of equation (26) for $r>1.25$,
is due to the initial outward migration
of the solid.

\section {Summary and Discussion}

We have studied the migration of solids
induced by gas drag and pressure gradients in a non-uniform
nebula. We considered a rotating, isothermal and
turbulence-free gaseous disk with an azimuthally symmetric
density enhancement on its midplane and a solar type star at its 
center. The results of our numerical simulations of the motions
of solids ranging from micron-sized particles to meter-sized
objects indicated that solids tend to migrate toward the
locations of local maximum pressure. In our model
nebula, such locations correspond to places where the
density of the gas is locally enhanced. Our results
indicate that depending on the physical properties
of solids, gas drag induced migration can be quite rapid.
Such rapid migrations can increase the rates of collision
and coalescence of particles and enhance their growth-rates
to larger bodies. The results of the extension of this study 
to include mass-growth
and collisional coagulation are currently under preparation for
publication.

It is necessary to emphasize that the assumptions of an isothermal
and turbulence-free nebula were made merely to focus attention 
on the sole effect of gas drag on the rate of migration of solids.
Also, in this study, we focused our attention on the motion of
objects without considering their mutual interactions.
In a realistic scenario, one has to allow for the changes of the
gas temperature on both radial and vertical directions and
also take turbulence and collision into consideration. It is expected that
by including turbulence, in particular in the vertical direction,
the rate of approach of particles to the midplane of the
nebula may decrease. It would then be of utmost value to investigate to 
what degree the combined effect of turbulence and gas drag induced
rapid migration will enhance the rate of the formation of
larger bodies. Such studies are currently underway.

\acknowledgments

This work is partially supported by the NASA Origins of the Solar 
System Program under Grant NAG5-10547 and by the NASA Astrobiology
Institute under Cooperative Agreement NCC2-1056.

\clearpage

\begin{figure}
\caption{$\qquad\qquad\qquad\qquad\qquad${\bf See Figure1.gif}
\label{fig1}}
\end{figure}

\clearpage

\begin{figure}
\plotone{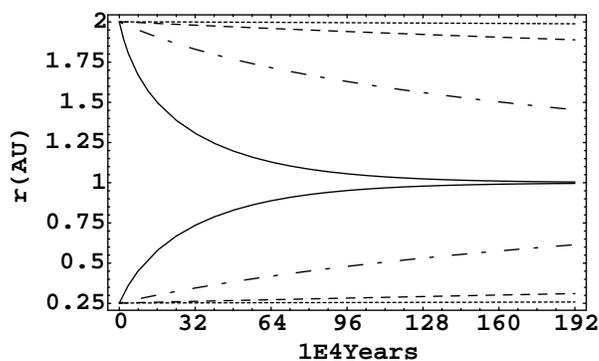}
\caption{Inward and outward radial migrations of small solids
from (2,0.2) AU and (0.25, 0.025) AU to (1,0) AU, respectively. 
The dotted line represents the migration of a micron-sized particle, the
dashed line is that of a 10 micron object, the dash-dotted
line corresponds to the radial migration of an object with a 
radius equal to 100 micron
and the solid line represents the migration of a 
millimeter-sized particle. The density of all particles is
equal to 2 g/cm$^3$. The physical properties
of the gas are similar to those of Figure 1.
\label{fig2}}
\end{figure}

\clearpage

\begin{figure}
\plotone{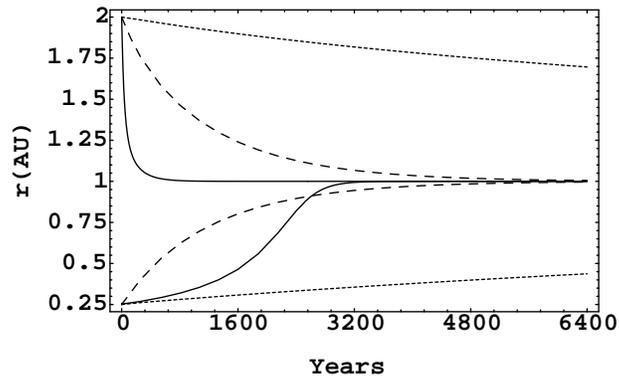}
\caption{Radial migration of centimeter-sized (dotted line),
10 centimeter-sized (dashed line) and meter-sized (solid line)
objects. The physical properties of the gas and the solids,
and the initial positions of the objects 
are similar to those of Figure 2. As shown here, meter-sized
objects show the most rapid radial migration.
\label{fig3}}
\end{figure}

\clearpage

\begin{figure}
\plotone{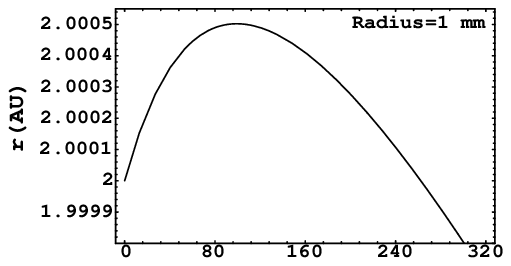}
\plotone{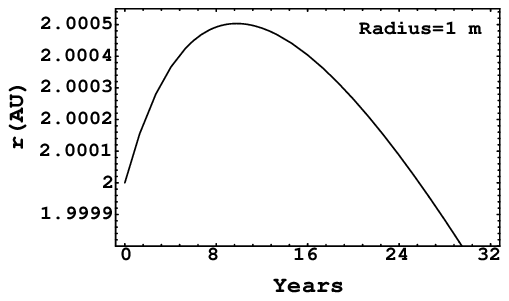}
\vskip -2in
\caption{Graphs of the initial outward migrations of 
a millimeter-size particle (top) and a centimeter-sized object (bottom).
Objects were initially placed at (2, 0.2) AU. The physical
properties of the gas and the solids are similar to 
those of Figure 2. Note different scales on the time axes.
\label{fig4}}
\end{figure}

\clearpage

\begin{figure}
\caption{$\qquad\qquad\qquad\qquad\qquad${\bf See Figure5.gif}
\label{fig5}}
\end{figure}

\clearpage

\begin{figure}
\plotone{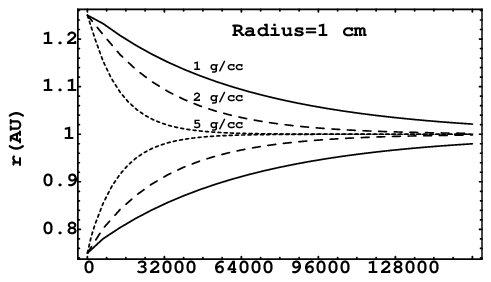}
\plotone{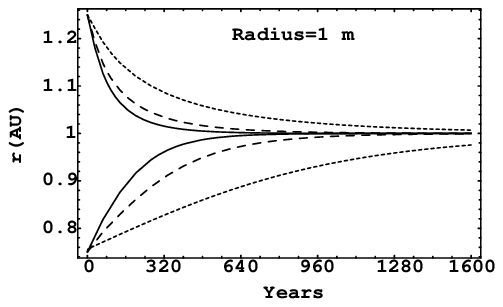}
\vskip -2in
\caption{Radial migrations of a centimeter-sized particle and
a meter-sized solid for different values of the solid's density.
The physical properties of the gas are similar to those of Figure 1.
As mentioned in the text, centimeter-sized and smaller particles
undergo more rapid radial migrations by increasing their densities.
The rate of radial migration of a meter-sized or larger object,
on the other hand, decreases by increasing its density.
Note different scales on time axes.
\label{fig6}}
\end{figure}

\clearpage

\begin{figure}
\plottwo{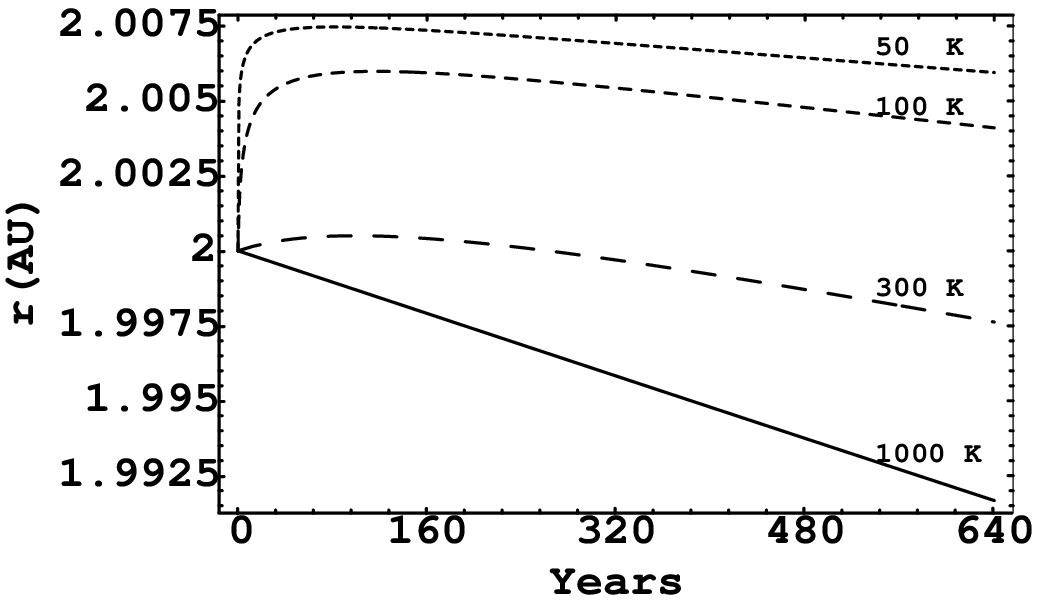}{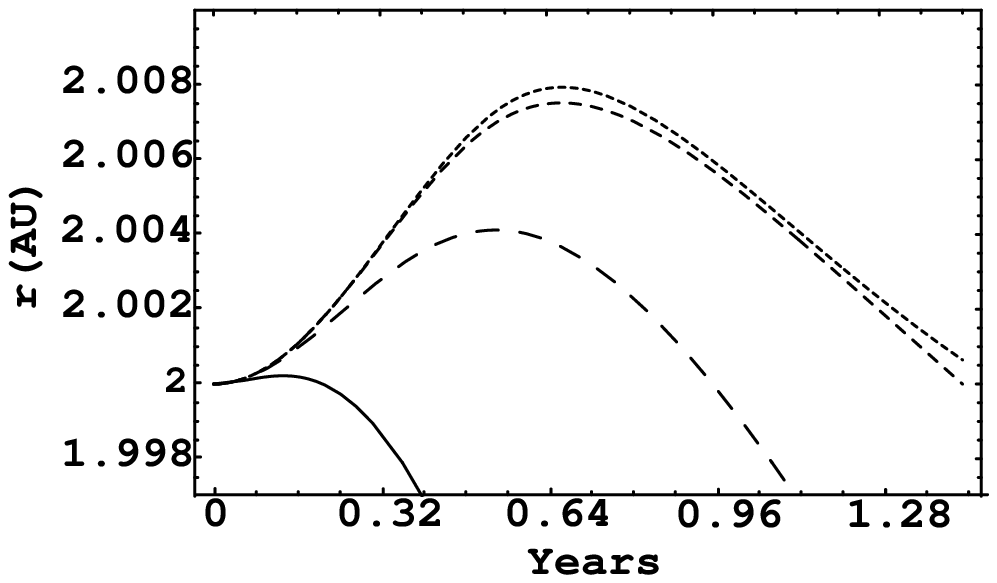}
\end{figure}
\begin{figure}
\plottwo{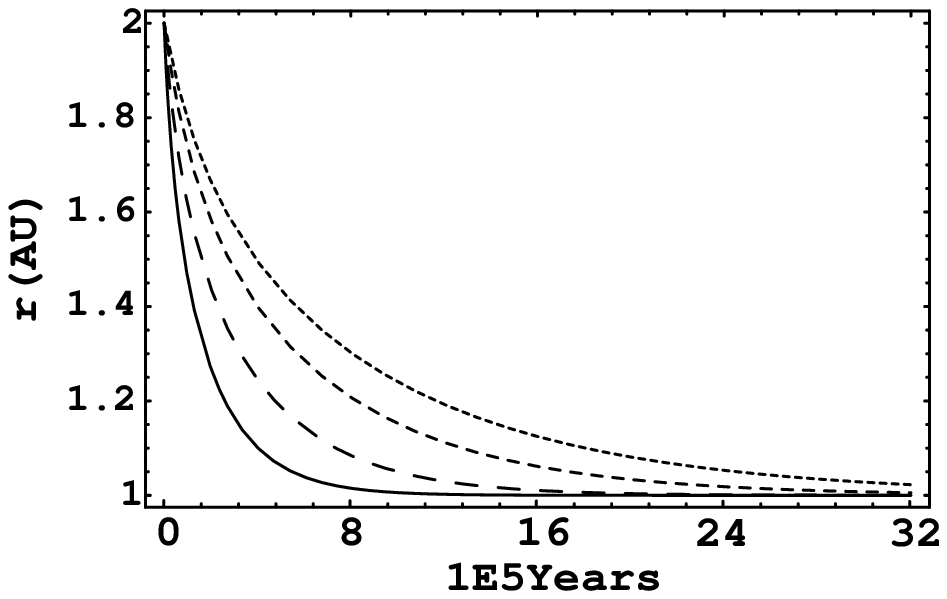}{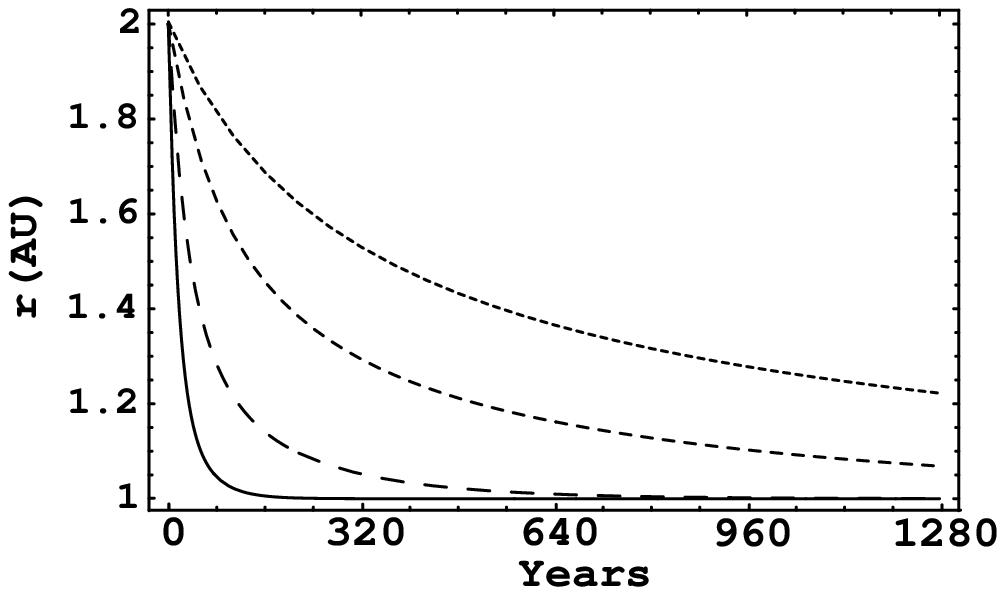}
\caption{Graphs of the radial migration of a millimeter-sized solid 
(left column) and a meter-sized object (right column) 
for different values of the
gas temperature. The values of ${\rho_0},\,\beta$ and $r_m$
are similar to those of Figure 1 and the density of all objects is
equal to 2 g cm$^{-3}$. As seen from this figure, the rate of radial 
migration increases by increasing the temperature. 
Note different scales on vertical and horizontal axes.
\label{fig7}}
\end{figure}

\clearpage

\begin{figure}
\plotone{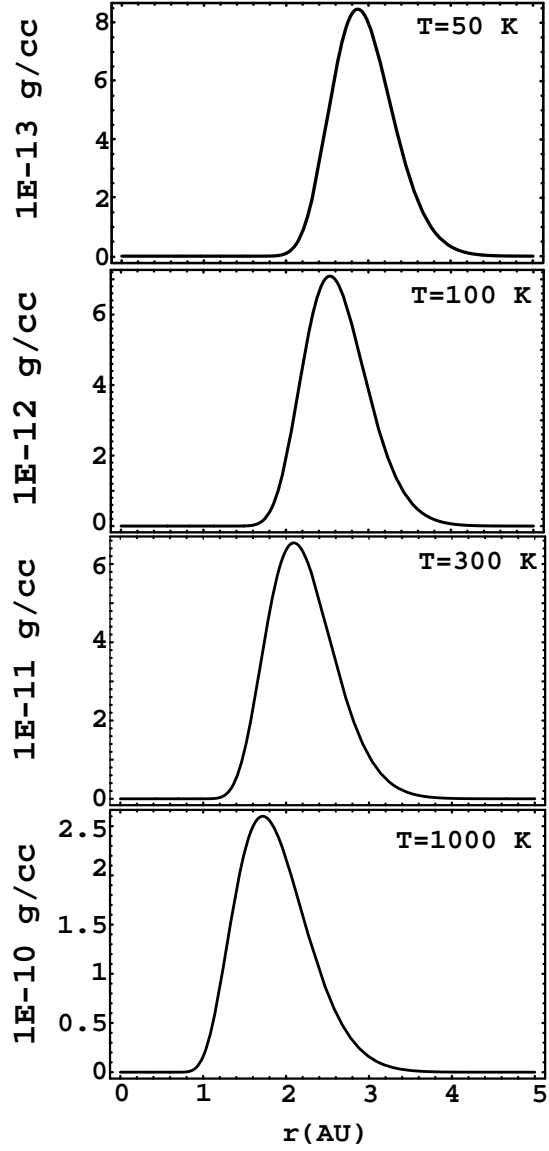}
\vskip -10in
\caption{Radial distribution of the gas on a plane parallel
to the midplane at $z=0.2$ AU. 
The values of ${\rho_0},\,\beta$ and $r_m$
are similar to those of Figure 1.
As shown here, increasing the gas temperature means that 
the value of the local maximum density increases and
its radial location approaches smaller distances.
Note different scales on vertical axes.
\label{fig8}}
\end{figure}

\clearpage

\begin{figure}
\plotone{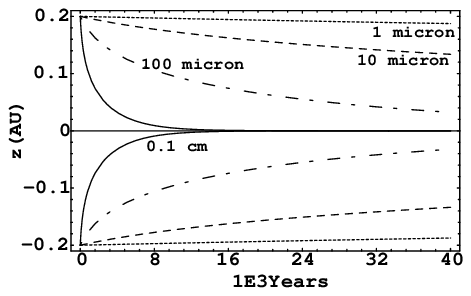}
\plotone{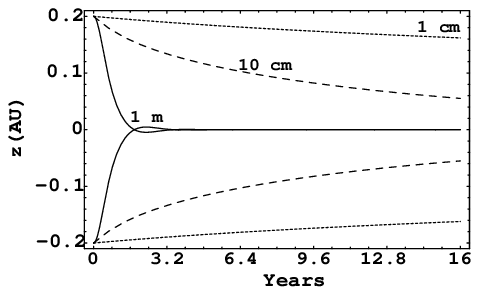}
\vskip -2in
\caption{Vertical migration of solids. The physical properties
of the nebula are identical to those of Figure 1.
Each particle was initially at a height equal to
one-tenth of its radial distance.
Note the different scales on the time axes.
\label{fig9}}
\end{figure}

\clearpage

\begin{figure}
\plotone{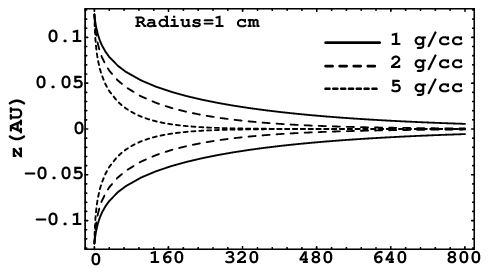}
\plotone{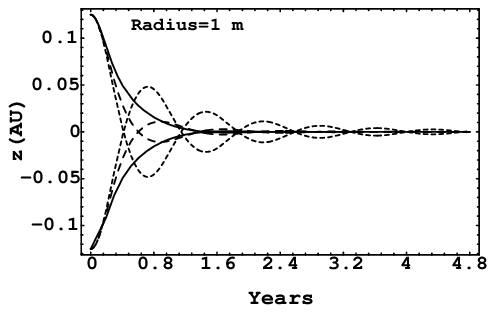}
\vskip -2in
\caption{Vertical motion of a centimeter-sized particle
and a meter-sized object in the vicinity of the midplane,
for different values of their densities. 
The initial positions of each
particle is (1.25, $\pm$0.125) AU.
The physical properties of the 
nebula are similar to those of Figure 1. 
As shown here, the motion of a 1 cm solid in the vicinity
of the midplane is overdamped while a 1 m object
undergoes an underdamped oscillatory motion. The rates of 
the initial descent of both solids increase by increasing their
densities. Note different scales on the time axes.
\label{fig10}}
\end{figure}

\clearpage

\begin{figure}
\caption{$\qquad\qquad\qquad\qquad\qquad${\bf See Figure11.gif}
\label{fig11}}
\end{figure}

\clearpage

\begin{figure}
\plottwo{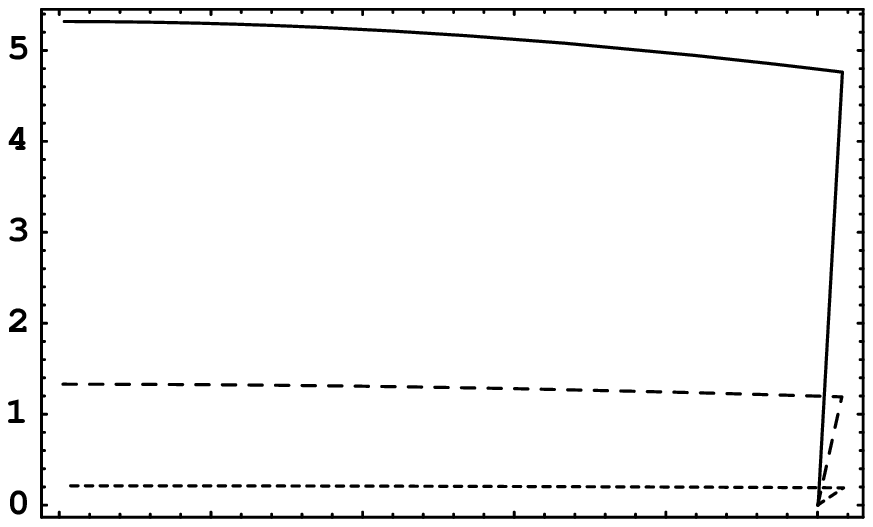}{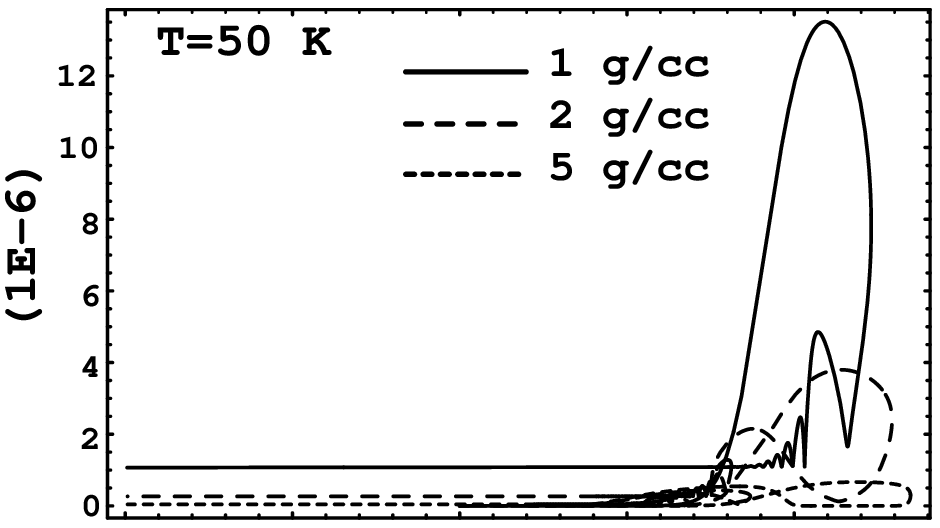}
\end{figure}
\begin{figure}
\plottwo{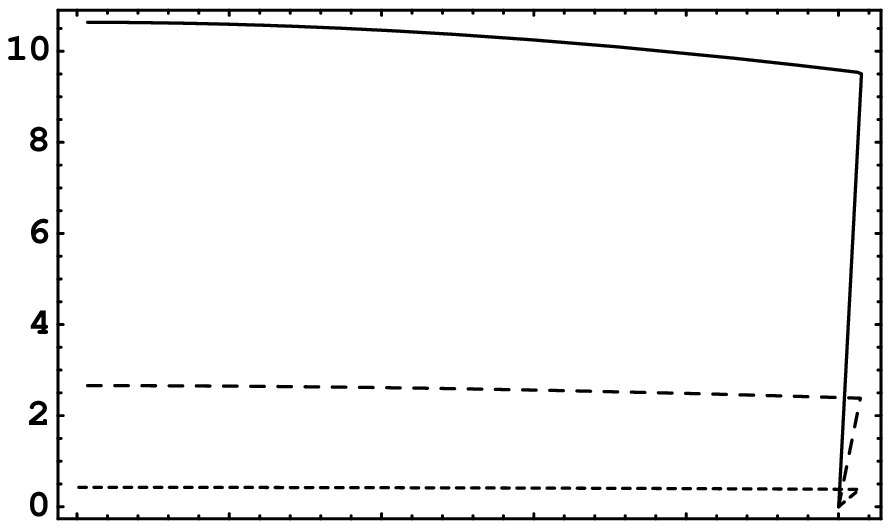}{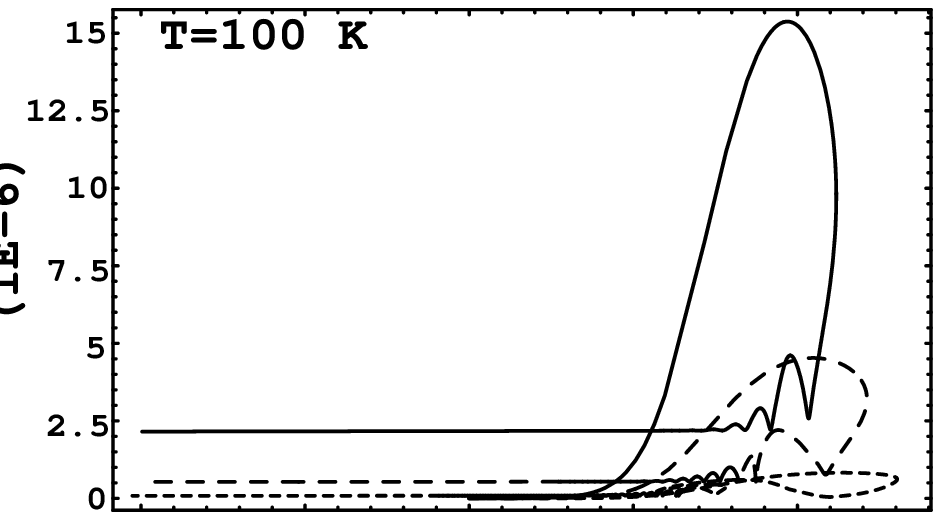}
\end{figure}
\begin{figure}
\plottwo{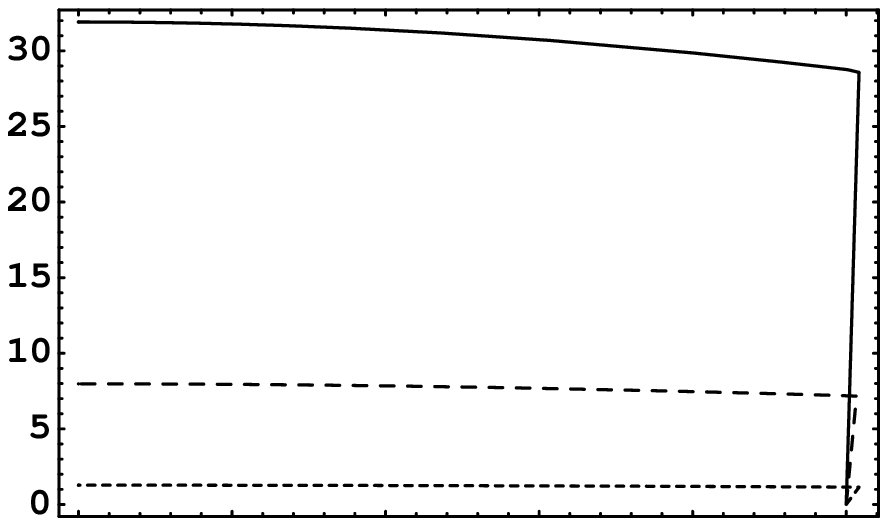}{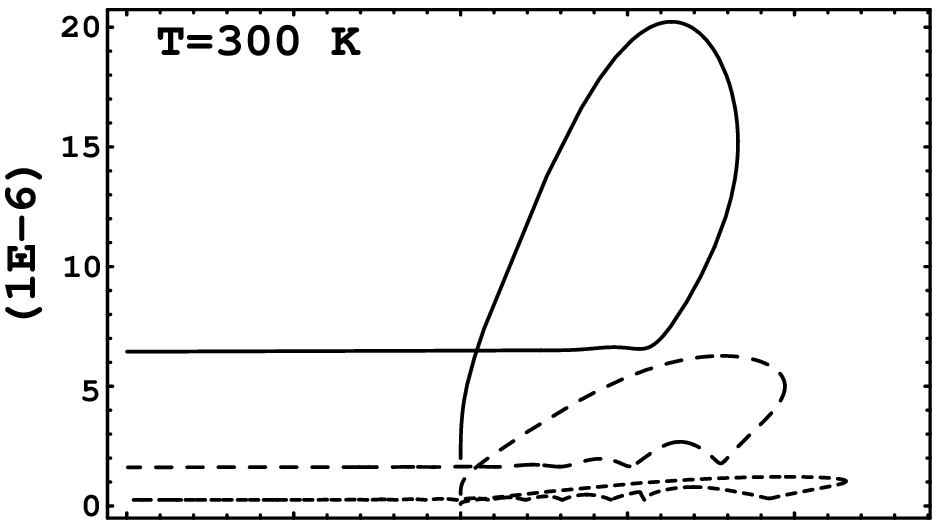}
\end{figure}
\begin{figure}
\plottwo{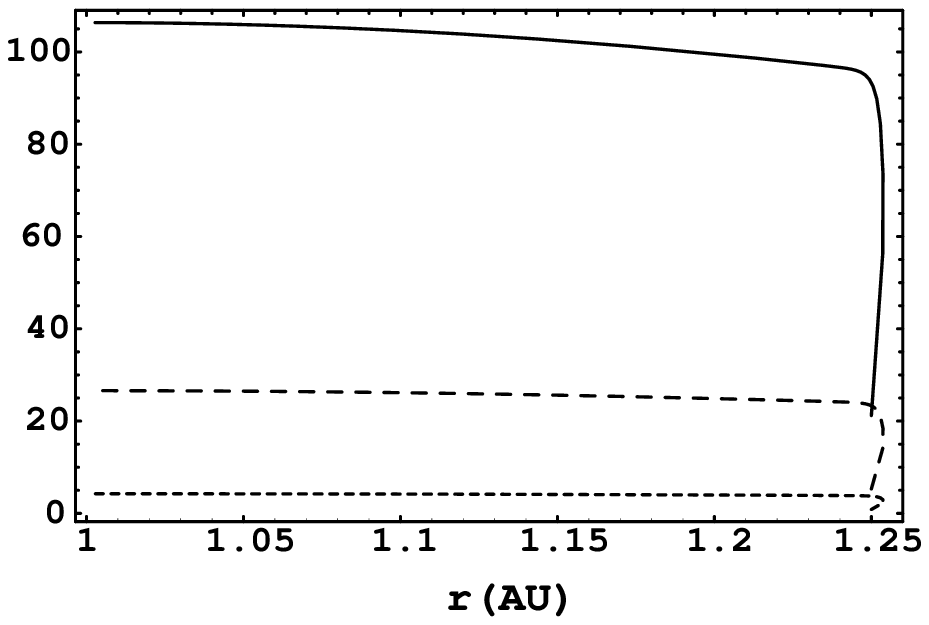}{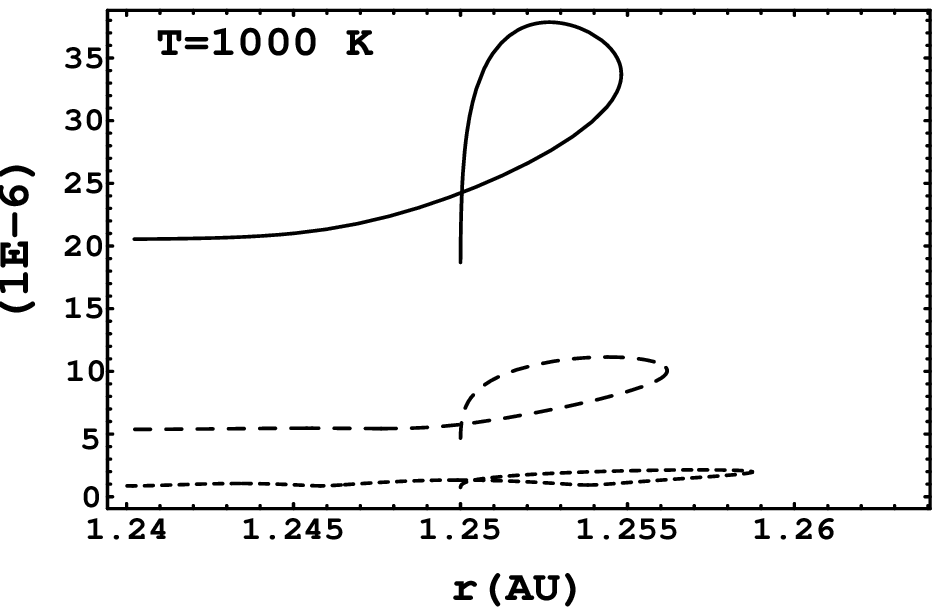}
\vskip -1in
\caption{Graphs of ${{\hat r}^{-3}}-{\hat \omega}^2$ versus $\hat r$
for a 1 cm particle (left column) and a meter-sized object
(right column). The objects were initially at (1.25,0.125) AU. 
The values of ${\rho_0},\, \beta$ and $r_m$
are similar to those of Figure 1.  
\label{fig12}}
\end{figure}

\clearpage

\begin{figure}
\plotone{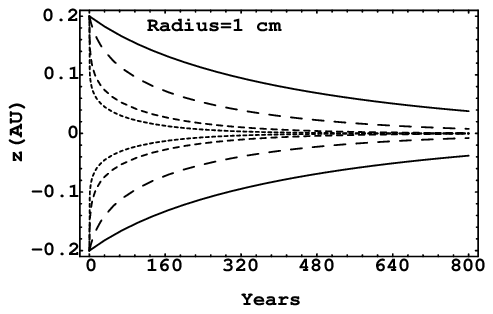}
\plotone{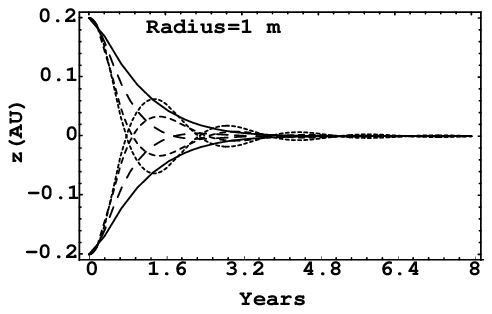}
\vskip -3in
\caption{Vertical migration of a centimeter-sized particle
and a meter-sized object 
with densities equal to 2 g cm$^{-3}$ for different
values of the gas temperature.
The values of ${\rho_0},\, \beta$ and $r_m$
are similar to those of Figure 1.  
Both particles were initially at
(2,$\pm$0.2) AU. Note different scales on time axes.
\label{fig13}}
\end{figure}

\clearpage

\begin{figure}
\plotone{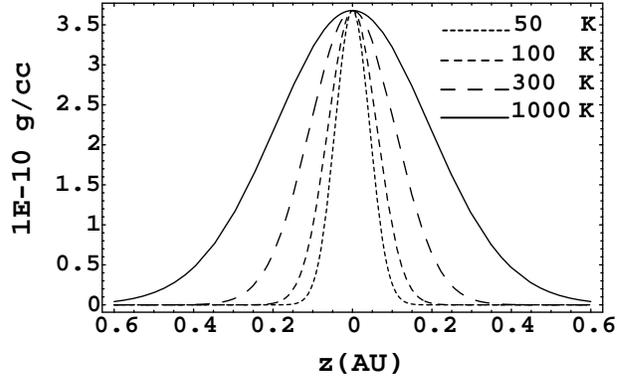}
\vskip -4in
\caption{Graph of the vertical distribution of gas at $r=2$ AU
for different values of temperature.
\label{fig14}}
\end{figure}

\clearpage

\begin{figure}
\caption{$\qquad\qquad\qquad\qquad\qquad${\bf See Figure15.gif}
\label{fig15}}
\end{figure}

\end{document}